# Experimental determination of the orientation of tilted magnetic anisotropy by measuring Hall voltage

Hasan Pişkin*, Erdem Demirci, Mustafa Öztürk and Numan Akdoğan

Gebze Technical University, Department of Physics, 41400 Gebze, Kocaeli, Turkey

***Corresponding author:** hpiskin@gtu.edu.tr

**Abstract**

The orientation of the tilted magnetic anisotropy has crucial importance in many spintronic devices. However, it is very challenging to determine it especially in very small structures produced by lithography. Here, we propose a new experimental method to directly and accurately measure both the polar and azimuthal angles of a tilted magnetic anisotropy. By using the proposed experimental method, we have successfully determined the out-of-plane and in-plane angles of the tilted magnetic anisotropy in a micro-structured multilayer thin film. The orientation of the tilted magnetic anisotropy in the sample has also been confirmed by the theoretical simulations proving the accuracy of the method.

## I. INTRODUCTION

Ultra-thin magnetic films with tilted out-of-plane magnetic anisotropy (TMA) have recently attracted considerable attention due to their key role in technological applications [1–5]. The use of TMA layer as a tilted (or canted) spin polarizer in spin-transfer torque (STT) based magnetic tunnel junctions has enabled a faster switching of magnetization direction and lower power consumption [6–8]. The STT structure has also been used as a more efficient spin transfer oscillator (STO) [7–9]. In addition, the presence of TMA in spin-orbit torque (SOT) devices allows the switching of the magnetization direction without an external magnetic field [2–4,10–13]. Therefore, the precise determination of the direction of the TMA in all the above-mentioned studies has a vital impact on the efficiency of the devices.

Although there are several direct and indirect experimental methods to determine the orientation of TMA, each of these methods has either weaknesses or difficulties. While torque magnetometry [14–16], generalized magneto-optical ellipsometry as vector magnetometry

[17,18], anisotropic interface magnetoresistance (AIMR) measurements [19], vibrating sample magnetometry (VSM), and scanning Kerr microscopy [20] can be mentioned as direct methods, ferromagnetic resonance [21], and Brillouin light scattering [22] are examples of the indirect methods. Among them, some can only determine the polar angle of the magnetic anisotropy [2,13,19], and some require a well-defined energy expression with a good determination of the magnetic anisotropy constants [21-22]. In particular, microfabricated devices often encounter limitations due to the alignment and sample size requirements of the above-mentioned methods.

In this work, we propose a very simple experimental method that can accurately determine both the polar and azimuthal angles of tilted magnetic anisotropy by measuring the Hall voltage. The proposed method is especially useful for micro devices due to the advantage of Hall effects which are very sensitive to the magnetization directions. With this motivation, we have fabricated a Hall-bar geometry consists a multilayer structure of Pt(4)/[Co(0.5)/Pt(0.5)]×5 /IrMn(8)/Pt(3) (nm) thin film obliquely deposited on a Si/SiO$_2$ (500 nm) substrate. The polar and azimuthal angles of the tilted magnetic anisotropy in the sample have been successfully determined by using the proposed experimental method. Furthermore, the experimentally obtained Hall voltages have been theoretically simulated by using magnetic free energy. The same polar and azimuthal angles of the magnetic anisotropy have been found from the theoretical calculations. The theoretically simulated Hall voltages also perfectly fit the experimental Hall signals at low in-plane magnetic fields. However, compared to the experiments, higher Hall voltages have been obtained from the simulations for certain in-plane angles of high magnetic fields. This behaviour has been explained by the decomposition of magnetic anisotropy into dynamic (rotatable) and static components.

## II. THEORETICAL BACKGROUND

In the electronic transport phenomena of a ferromagnetic (FM) thin film, the Hall voltage ($V_{Hall}$) can be expressed with three contributions from ordinary (OHE), anomalous (AHE) and planar (PHE) Hall effects [23,24]:

$$V_{Hall} = V_{OHE} + V_{AHE} + V_{PHE} \quad (1)$$

Although the AHE and PHE voltages originate from the spin-orbit coupling in FM materials [5,25,26], the OHE voltage can be observed in any conductor due to the Lorentz force exerting on electrons when an external magnetic field applied [27]. Since the OHE contribution is very small compared to the AHE and PHE at relatively low magnetic fields, the Eq. (1) can be written as follows [23];

$$V_{Hall} = V_{AHE} + V_{PHE},  \quad (2)$$

Thus, for a magnetic system with a tilted magnetic anisotropy as presented in Fig. 1, the transverse voltage is given by following expression with an applied current along the x-axis;

$$V_{Hall} = \frac{i_x}{2}\Delta R_{AHE} \cos\theta_M + \frac{i_x}{2}\Delta R_{PHE} \sin 2\varphi_M \sin^2\theta_M, \quad (3)$$

where the $\Delta R_{AHE}$ and $\Delta R_{PHE}$ are the resistance differences for the anomalous Hall and the planar Hall effects, respectively. The $\theta_M$ and $\varphi_M$ are the polar and azimuthal angles of the magnetization vector (**M**).

When the applied magnetic field (**H**) is zero, the magnetization is controlled by the effective magnetic anisotropy and the $\theta_K = \theta_M$ and $\varphi_K = \varphi_M$ conditions occur. Thus, Eq. (3) can be expressed as the remanence of Hall voltage ($V_R^{Hall}$);

$$V_R^{Hall} = \frac{i_x}{2}\Delta R_{AHE} \cos\theta_K + \frac{i_x}{2}\Delta R_{PHE} \sin 2\varphi_K \sin^2\theta_K \quad (4)$$

where the $\theta_K$ and $\varphi_K$ are the polar and azimuthal angles of the effective magnetic anisotropy as demonstrated in Fig. 1.

If the Eq. (4) is rearranged, the following function can be written:

$$\cos^{-1}\left(\frac{V_R^{Hall} - \frac{i_x}{2}\Delta R_{PHE} \sin 2\varphi_K \sin^2\theta_K}{\frac{i_x}{2}\Delta R_{AHE}}\right) - \theta_K = 0 \quad (5)$$

The $V_R^{Hall}$, $\Delta R_{AHE}$, and $\Delta R_{PHE}$ values in Eq. 5 can be obtained from the conventional anomalous and planar Hall measurements. The applied current ($i_x$) is also known from the experiment. If one of the $\varphi_K$ or $\theta_K$ angles is determined, the other one can be calculated by using Eq. 5.

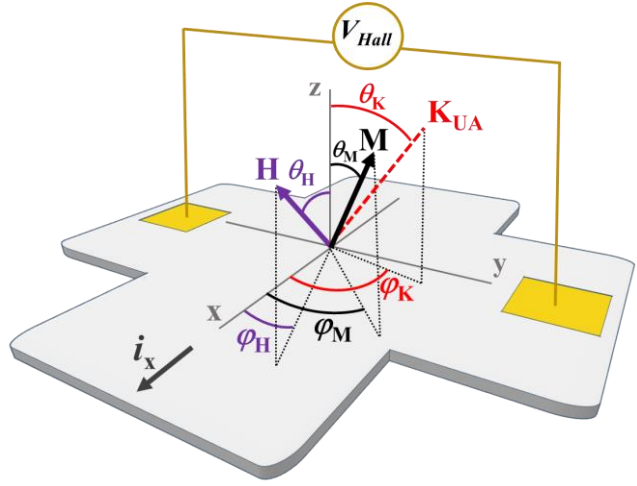

FIG. 1. A schematic representation of the coordinate system for the magnetization (**M**), the easy axis of the tilted magnetic anisotropy (**K$_{UA}$**) and the applied magnetic field (**H**). The polar angles are denoted by $\theta$ according to the z-axis and the azimuthal angles are shown by $\varphi$ according to the x-axis. A constant current of $i_x$ is applied along the x-axis and the $V_{Hall}$ is measured from contact pads along the y-axis of the cross geometry.

In order to obtain the $V_R^{Hall}$, $\Delta R_{AHE}$, and $\Delta R_{PHE}$ values from the experiment, we can consider a ferromagnetic film with a tilted magnetic anisotropy ($\theta_K \neq 0°$). A typical $V_{Hall}$ signal of such a system can be plotted as a function of the magnetic field applied along the z-axis, as shown in Fig. 2(a). If the magnetization is saturated along the z-axis, the $\theta_M$ in Fig. 1 will be equal to 0° or 180°. In this case, the second term in Eq. (3) disappears and $V_{Hall}$ equals to $V_{AHE}$. Since the first term is a function of $\cos\theta_M$, the $V_{Hall}$ reaches its maximum ($+V_S^{AHE}$) and minimum ($-V_S^{AHE}$) values for $\theta_M = 0°$ and $\theta_M = 180°$ conditions, respectively. If the voltage difference between $+V_S^{AHE}$ and $-V_S^{AHE}$ is divided by the applied current $i_x$, the $\Delta R_{AHE}$ can be obtained from the $V_{Hall}$ signal as shown in Fig. 2(a). It is noteworthy that the hysteresis curve presented in Fig. 2(a) can be shifted left or right side in exchange-biased systems. In such cases, there may be single or two different $V_R^{Hall}$ values depending on the magnitude of the exchange-bias. If there are two $V_R^{Hall}$ values, they can be used to calculate the direction of magnetization in either direction relative to the z-axis.

Furthermore, if the magnetization can be saturated in the film plane by an applied in-plane magnetic field ($\theta_H = \theta_M = 90°$ and $\varphi_H = \varphi_M$ conditions in Fig. 1), the first term in Eq. (3) becomes zero and the Hall voltage contains only the PHE signal. Fig. 2(b) presents a typical Hall voltage behaviour as a function of in-plane angle of the magnetic field ($\varphi_H$). Due to the $\sin 2\varphi_M$ dependence of the PHE signal, the $V_{Hall}$ becomes maximum ($+V_S^{PHE}$) for $\varphi_H = 45°$ and 225° angles, and it goes to its minimum value ($-V_S^{PHE}$) for $\varphi_H = 135°$ and 315° angles. The voltage difference between $+V_S^{PHE}$ and $-V_S^{PHE}$ in Fig. 2(b) gives the $\Delta R_{PHE}$ when it is divided by the applied current $i_x$.

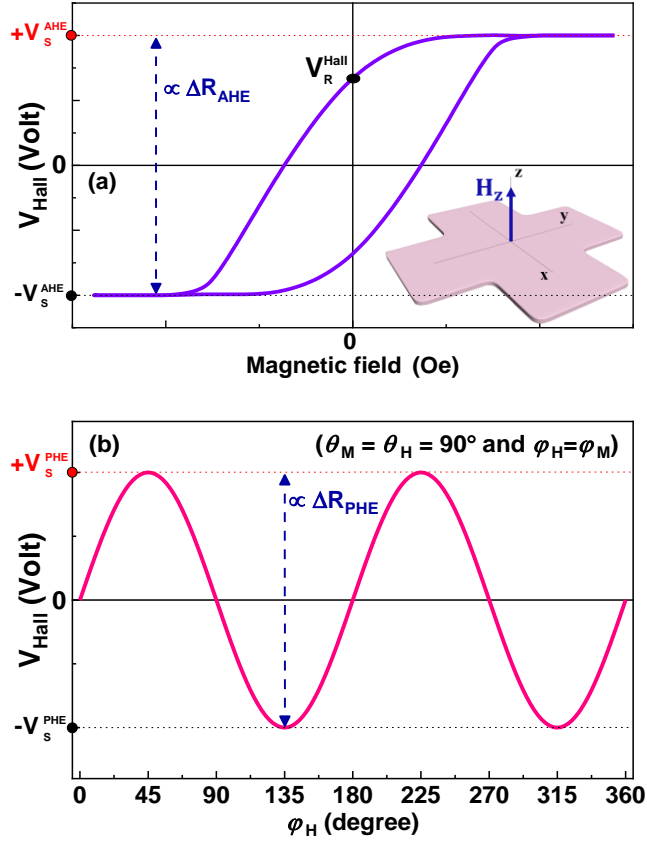

FIG. 2. The typical Hall voltage signals for two different measurement geometries. (a) The Hall voltage as a function of the magnetic field applied along the z-axis plotted for a conventional AHE geometry. (b) The Hall signal as a function of $\varphi_H$ for a magnetic field applied in the film plane ($\theta_H = \theta_M = 90°$). It should be noted that the magnitude of the magnetic field should be sufficient to saturate the magnetization in the film plane to provide a conventional PHE signal.

As a next step, we propose a simple method to determine the azimuthal angle of magnetic anisotropy ($\varphi_K$) from Hall measurements. For this purpose, we consider a sample with the tilted magnetic anisotropy angles of $\theta_K = \varphi_K = 45°$, as schematically shown in the inset of Fig. 3. When the magnetic field is zero. the magnetization would be along the easy axis with $\theta_M = \varphi_M = 45°$ conditions. If a relatively weak magnetic field is applied in the film plane ($\theta_H = 90°$), the magnetization cannot overcome the hard axis barrier and it can be slightly manipulated around the easy axis. This results in a change in the Hall voltage as a function of $\varphi_H$ due to the changing of $\theta_M$ and $\varphi_M$ in Eq. (3). A typical $\varphi_H$ dependence of the Hall voltage is given in Fig. 3 for a system with the tilted magnetic anisotropy ($\theta_K = \varphi_K = 45°$). The details of the calculation are given in Supplementary Material. Two limit conditions can be discussed here for the Hall voltage as a function of $\varphi_H$. (i) When a weak in-plane magnetic field is applied for varying $\varphi_H$ angles between 0° and 90°, the polar angle of magnetization ($\theta_M$) slightly increases. The small increase in $\theta_M$ causes a detectable decrease in $V_{Hall}$ due to its dependence on $\theta_M$ in Eq. (3). It is worth emphasizing that the bigger $\theta_M$ results in an increase in the second term of Eq. (3). However, the total behaviour of the Hall voltage is governed by the change in the first term, since the $\Delta R_{AHE}$ is bigger than $\Delta R_{PHE}$ for ferromagnetic materials [24,28,29]. (ii) If the same amount of in-plane magnetic field is applied between $\varphi_H$ angles of 180° and 270°, the $\theta_M$ decreases slightly and gets closer to the z-axis. Thus, the Hall voltage has a peak in that region of the $\varphi_H$ angle. A maximum value occurs in $V_{Hall}$ when the in-plane magnetic field is applied at $\varphi_H=225°$ as shown in Fig. 3. This angle can also be defined as $\varphi_H = \varphi_K+180°$, since the $\varphi_K$ is given as 45°. By using that relation, the unknown orientation of $\varphi_K$ in tilted magnetic systems can be determined from the angle where the maximum Hall voltage is recorded as a function of $\varphi_H$. Thus, the polar angle of magnetic anisotropy ($\theta_K$) can be calculated by using Eq. (5).

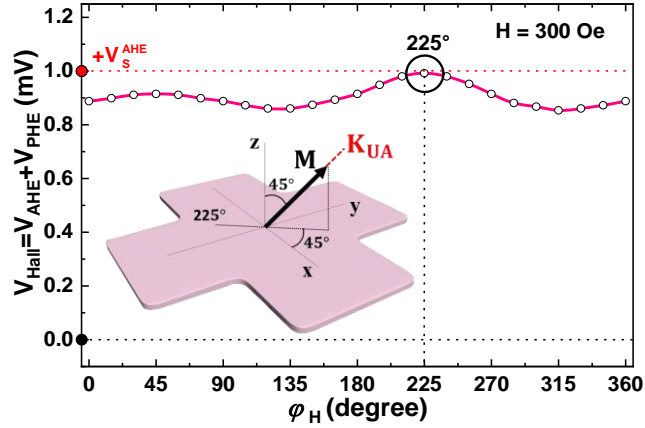

FIG. 3. The $V_{Hall}$ signal has been calculated as a function of $\varphi_H$ angle for a fixed in-plane magnetic field of 300 Oe. The coordinate system used in the calculations is shown schematically in the inset.

## III. EXPERIMENTAL RESULTS

The Hall-bar geometry with dimensions of 80μm×80μm has been fabricated by optical lithography on a Si/SiO$_2$ (500 nm) substrate. Then, a multi-layer structure of Pt(4)/[Co(0.5)/Pt(0.5)]×5/IrMn(8)/Pt(3) (nm) has been obliquely deposited at room temperature by magnetron sputtering with a base pressure of 1.5×10$^{-9}$ mbar. A dc current of 1 mA has been applied during the experiments by using a Keithley 2400 sourcemeter. The Hall voltages have been recorded at room temperature by a Keithley 2002 multimeter.

Fig. 4(a) presents the measured Hall voltage ($V_{Hall}$) as a function of magnetic field applied perpendicular to the film plane ($\theta_H$ = 0° and 180°). The data indicates a typical magnetic hysteresis curve with an exchange bias of 20 Oe. The saturation voltages ($\pm V_S^{AHE}$) of ±0.575 mV have been obtained when the sample is saturated along the z-axis. This gives the $\Delta R_{AHE}$ value of 1.150 Ω which will be used in Eq. (5). In addition, the $V_R^{Hall}$ = 0.557 mV and $-V_R^{Hall}$ = -0.555 mV values have been found for the +z and -z directions, respectively. This slight difference in between is due to the presence of an exchange bias in the sample. Among them, the positive remanence Hall voltage ($V_R^{Hall}$) will be used in Eq. 5 to calculate $\theta_K$ along the +z direction.

We have also carried out Hall experiments as a function of the in-plane magnetic field of 500 Oe for varying $\varphi_H$ angles between 0° and 360°. The Hall voltage ($V_{Hall}$) presented in Fig. 4(b) exhibits a peak at $\varphi_H$ = 39°. This peak shows that the easy axis of tilted magnetic anisotropy lies in $\varphi_K$ = 39° + 180° = 219°, as described in the theoretical background. Furthermore, we have repeated the same experiment by applying higher magnetic fields up to 5500 Oe to fully saturate the magnetization along the film plane. This measurement is important to obtain the $\Delta R_{PHE}$ value which is required to calculate $\theta_K$ in Eq. (5). The data presented in Fig. 4(c) indicates that even the magnetic field of 5500 Oe (the limit of the used experimental setup) was not enough to saturate the magnetization in the film plane. Thus, unlike the Fig. 2(b), an asymmetric signal of $V_{Hall}$ has been observed above the zero voltage. However, the $\Delta R_{PHE}$ value of 0.2 Ω can still be estimated from the maximum and minimum of the $V_{Hall}$ taken at 5500 Oe. The theoretical simulations presented in the following part support this data. Finally, if the $V_R^{Hall}$ = 0.557 mV, $\Delta R_{AHE}$ = 1.150 Ω, $\Delta R_{PHE}$ = 0.20 Ω and $\varphi_K$ = 219° values obtained from the Hall experiments are used in the Eq. (5), the polar angle ($\theta_K$) of easy axis can be calculated as 24.4° for the studied sample system.

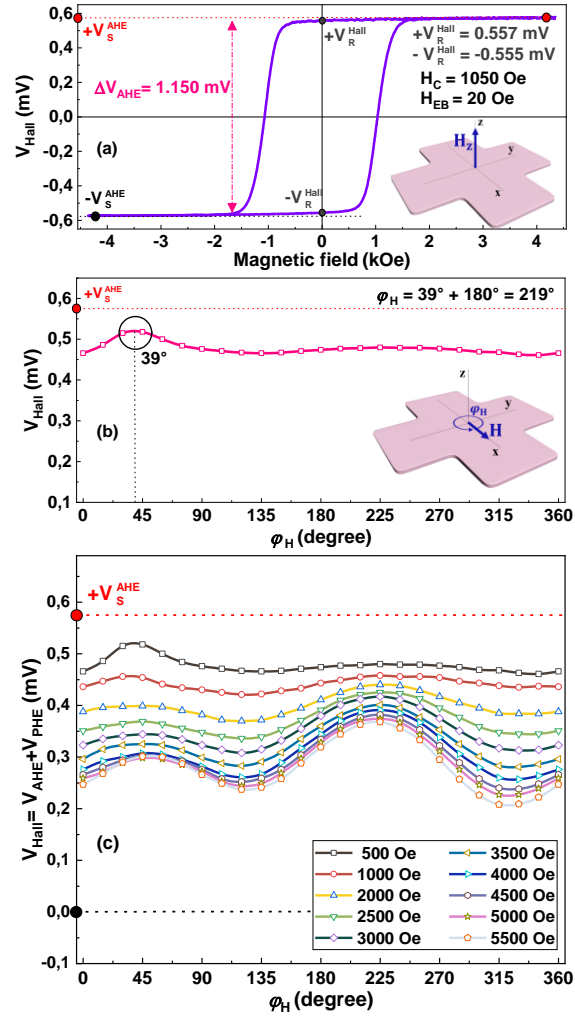

FIG. 4. The Hall voltages as a function of magnetic field (**H**) applied perpendicular (a) and parallel (b) to the film plane. The peak in figure (b), observed at 39°, gives the azimuthal angle of tilted magnetic anisotropy with the relation of $\varphi_K = 39° + 180° = 219°$, which has been explained in the theoretical background. (c) The Hall voltage recorded as a function of $\varphi_H$ for varying in-plane magnetic fields up to 5500 Oe. Solid lines are guide for the eye.

It is important to note that the $V_{Hall}$ shown in Fig. 4(c) never drops to below zero voltage. However, when the magnitude of the in-plane magnetic field is sufficient to overcome the hard axis barrier in a tilted magnetic anisotropy system, the magnetization direction should switch from one easy direction to another one for certain $\varphi_H$ angles. This would result in a negative Hall voltage in the experiments. This has been demonstrated in the Supplementary Material by using the traditional Stoner-Wohlfarth (SW) model for a tilted magnetic anisotropy system. Since this is not the case in the studied sample, the unusual behaviour of the magnetization is explained by the presence of a rotatable magnetic anisotropy (RMA) in which the magnetic anisotropy axis (**K_UA**) can rotate towards the applied magnetic field [30–37]. Thus, the magnetization always stays in the +z region for varying $\varphi_H$ angles. The theoretical simulations of the experimental results support this conclusion.

## IV. THEORETICAL SIMULATION OF THE EXPERIMENTAL RESULTS

The experimentally observed Hall voltages ($V_{Hall}$) presented in Fig. 4 can be theoretically simulated by using Eq. (3), if the direction of magnetization ($\theta_M$ and $\varphi_M$) is known for each magnetic field step. This can be achieved by using the magnetic free energy of the system consists of a rotatable magnetic anisotropy, unidirectional anisotropy due to the exchange bias, and the Zeeman energy terms, respectively:

$$E = -K_{UA}\left(\frac{\mathbf{M}.\hat{\mathbf{u}}_{\mathbf{rot}}}{M}\right)^2 - J_{EB}\left(\frac{\mathbf{M}.\hat{\mathbf{u}}_\mathbf{0}}{M}\right) - \mathbf{M}.\mathbf{H}, \tag{6}$$

In Eq. (6), the orientation of the $K_{UA}$ is defined by the unit vector $\hat{\mathbf{u}}_{\mathbf{rot}}$ which is the magnetic field dependent [38], and the direction of the exchange bias is given by the unit vector $\hat{\mathbf{u}}_\mathbf{0}$. The magnetic field dependence of the rotatable uniaxial anisotropy can be expressed as follows;

$$\hat{\mathbf{u}}_{\mathbf{rot}} = u_x + u_y + u_z, \tag{7}$$

$$u_x = \sin(\theta_K + \Delta\theta_K)\cos(\varphi_K + \Delta\varphi_K), \tag{8}$$

$$u_y = \sin(\theta_K + \Delta\theta_K)\sin(\varphi_K + \Delta\varphi_K), \tag{9}$$

$$u_z = \cos(\theta_K + \Delta\theta_K), \tag{10}$$

where, the $\theta_K$ and $\varphi_K$ indicate the initial state of the effective magnetic anisotropy. The $\Delta\theta_K$ and $\Delta\varphi_K$ values depend on the magnetic field vectors $\mathbf{H}$ ($H_x + H_y + H_z$) and $\mathbf{H_K}$ ($H_{Kx} + H_{Ky} + H_{Kz}$). Thus, the $\theta_K + \Delta\theta_K$ and $\varphi_K + \Delta\varphi_K$ can be written as follows;

$$\theta_K + \Delta\theta_K = \cos^{-1}\left(\frac{H_{Kz}+\lambda_\theta.H_z}{H_K + \lambda_\theta.H}\right), \tag{11}$$

$$\varphi_K + \Delta\varphi_K = \tan^{-1}\left(\frac{H_{Ky} + \lambda_\varphi.H_y}{H_{Kx} + \lambda_\varphi.H_x}\right), \tag{12}$$

where the $\mathbf{H_K}$ is the magnetic field equivalence of the $K_{UA}$ which can be expressed by [39]:

$$H_K \equiv \frac{2K_{UA}}{M} \tag{13}$$

The $\lambda_\theta$ and $\lambda_\varphi$ terms in Eqs. (11) and (12) are restrictive parameters which define the magnetic field dependence of the magnetic anisotropy ($K_{UA}$) for its out-of-plane and in-plane components, respectively. If the $\lambda_\theta$ and $\lambda_\varphi$ were zero, the Eq. (6) would result in the traditional Stoner-Wohlfarth model due to the $\hat{u}_{rot}=\hat{u}_0$ condition. For this study, we have used $\lambda_\theta=0.15$ and $\lambda_\varphi=1$ which produce a change in the polar ($\Delta\theta_K$) and azimuthal ($\Delta\varphi_K$) angles of the effective magnetic anisotropy, due to the applied magnetic field.

Figs. 5(a) and (b) show the $\theta_M$ and $\varphi_M$ values calculated by using Eq. (6) as a function of $\varphi_H$ for different magnetic field values. It is important to mention that the experimentally obtained $\theta_K = 24.4°$ and $\varphi_K = 219°$ values have been used in Eq. (6) in order to define the initial condition ($\hat{u}_0$) of the system when the magnetic field is zero. The simulation results presented in Fig. 5(a) clearly indicates that the direction of the magnetization never goes to the negative z region. It always stays in the positive z region ($0° < \theta_M < 90°$) for varying $\varphi_H$ angles between 0° and 360°. This agrees with the experimental results presented in Fig. 4(c). In addition, it has been found that the magnitude of $K_{UA}$ depends on the in-plane magnetic field as shown in Fig. 5(c). It should be increased as a function of the magnitude of the magnetic field to fit the experimental results given in Fig. 4(c). This type of magnetic field dependence of the magnetic anisotropy has been previously reported in the literature [40,41]. Fan *et al.* explained the origin of this behaviour with a non-uniform magnetization in such systems. [41].

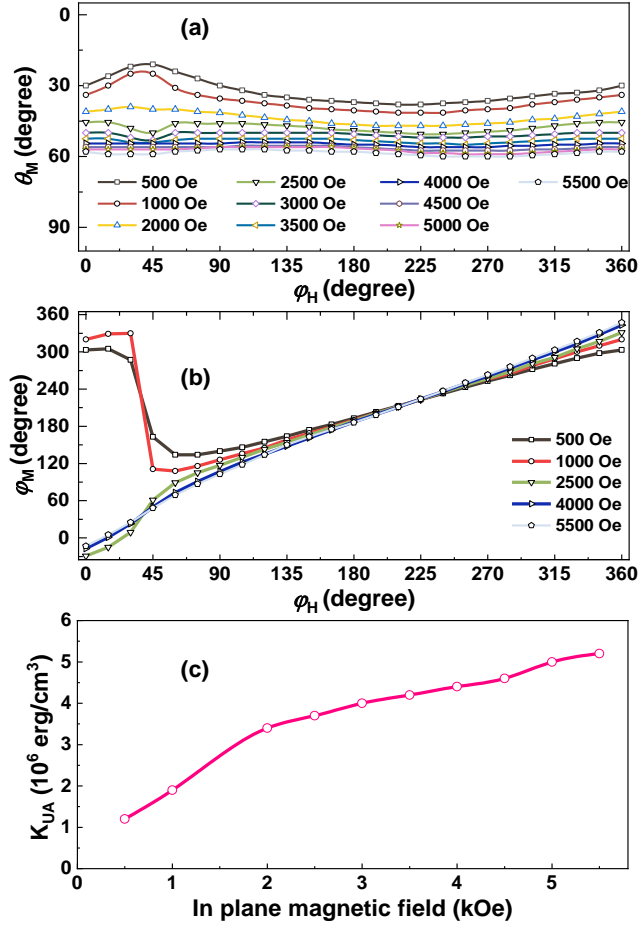

FIG. 5. The polar ($\theta_M$) (a) and the azimuthal ($\varphi_M$) (b) angles of magnetization as a function of $\varphi_H$. The data has been calculated for different magnetic field values by using Eq. (6). (c) The in-plane magnetic field dependence of the magnetic anisotropy ($K_{UA}$). The solid lines are guide for the eye.

Thus, the $V_{Hall}$ signals in Fig. 4(c) can be simulated as a function of $\varphi_H$ by inserting the calculated magnetization angles ($\theta_M$ and $\varphi_M$) and the experimentally obtained $\Delta R_{AHE} = 1.15 \ \Omega$ and $\Delta R_{PHE} = 0.2 \ \Omega$ values into the Eq. (3). Figure 6 presents the theoretically calculated Hall voltages plotted as a function of the in-plane magnetic field. In general, the simulated $V_{Hall}$ signal is compatible with the experimental results and supports the theoretical model proposed in this study. For low magnetic fields, the $V_{Hall}$ gives a peak at $\varphi_H = 39°$ and the shape of the signal fits very well to the experimental data shown in Fig. 4. However, for the higher magnetic fields, there is a considerable difference between the experiment and the simulation for the angle interval of $\varphi_H = 0°$ and $\varphi_H = 135°$, which is indicated as Region (I) in Fig. 6. In that region, the calculated Hall voltage is higher than the experimental results presented in Fig.

4(c). This difference has been investigated theoretically by applying combinations of RMA and SW models for the sample system under study.

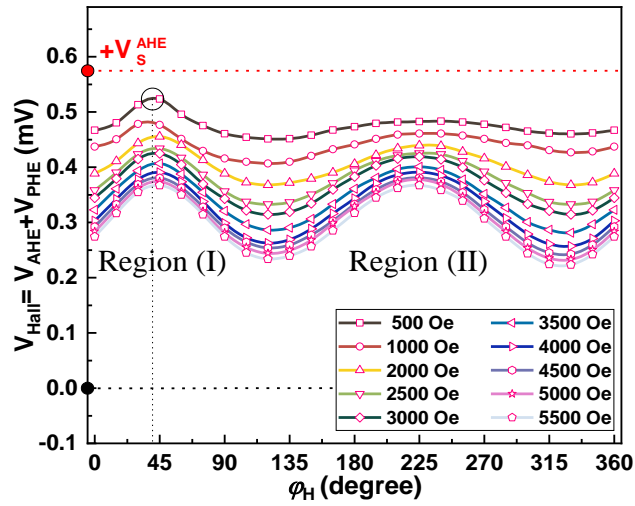

FIG. 6. The Hall voltages calculated by using Eq. (3) and RMA model as a function of $\varphi_H$ for varying in-plane magnetic fields up to 5500 Oe. The solid lines are guide for the eye.

Figure 7 presents the experimental and the simulated Hall voltages for the in-plane magnetic field of 4500 Oe. Although the RMA model always gives a positive $V_{Hall}$ for any in-plane angle of the magnetic field, the SW model exhibits a negative voltage in Region (I) since the applied magnetic field is enough to overcome the hard axis barrier as it has been discussed in Supplementary Material. The experimental data can only be perfectly simulated by using the linear combination of Hall voltages produced by 90% RMA and 10% SW models in the calculations. This indicates that approximately 90% of the magnetic moments in the sample are driven by the RMA model and the rest by the SW model. Such decomposition of magnetic anisotropy into static and dynamic (rotatable) components has been previously reported and supports this result [36].

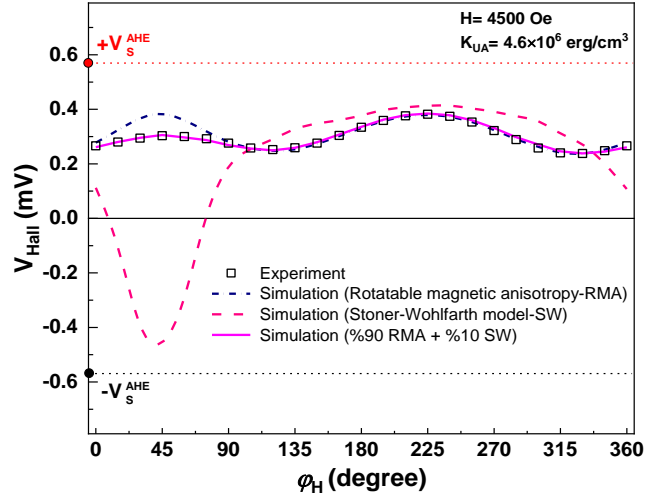

FIG. 7. The experimental (open squares) and simulated Hall voltages for the in-plane magnetic field of 4500 Oe. While the simulation with the SW model (dashed line) exhibits negative Hall voltage for certain $\varphi_H$ angles, the RMA model (dashed dot line) always results in a positive voltage. The solid line shows the linear combination of the Hall voltages with % 90 RMA and % 10 SW contributions, which perfectly fits the experiment.

## V. CONCLUSION

In conclusion, we have experimentally demonstrated that the tilted out of plane orientation of the magnetic anisotropy can be accurately determined by measuring the Hall voltage. The theoretical calculations proved that the proposed experimental method provides both the polar and azimuthal angles of magnetic anisotropy without the need for simulation. A multilayer structure of Pt(4)/[Co(0.5)/Pt(0.5)]×5/IrMn(8)/Pt(3) (nm) thin film has been fabricated to carry out the intended measurements. The experimental data and the theoretical calculations indicate that the magnetization behaviour of the sample has been dominated by a rotatable magnetic anisotropy which allows the tilted magnetic anisotropy of the system to rotate towards the applied magnetic field. Thus, the developed method gives information not only about the orientation of anisotropy but also about its character and magnitude.

## ACKNOWLEDGEMENTS

We would like to thank Sinan Kazan, Yıldırhan Öner, and Bulat Rameev for fruitful discussions.